# Holographic Air-quality Monitor (HAM)

Nicholas Bravo-Frank[1,2], Lei Feng[2], and Jiarong Hong[1,2,3]

1: Department of Electrical and Computer Engineering, University of Minnesota
2: Saint Anthony Falls Laboratory, University of Minnesota
3: Department of Mechanical Engineering, University of Minnesota

E-mail: jhong@umn.edu

**Abstract**

We introduce the holographic air-quality monitor (HAM) system, uniquely tailored for monitoring large particulate matter (PM) over 10 µm in diameter—particles critical for disease transmission and public health but overlooked by most commercial PM sensors. The HAM system utilizes a lensless digital inline holography (DIH) sensor combined with a deep learning model, enabling real-time detection of PMs, with greater than 97% true positive rate at less than 0.6% false positive rate, and analysis of PMs by size and morphology at a sampling rate of 26 liters per minute (LPM), for a wide range of particle concentrations up to 4000 particles/L. Such throughput not only significantly outperforms traditional imaging-based sensors but also rivals some lower-fidelity, non-imaging sensors. Additionally, the HAM system is equipped with additional sensors for smaller PMs and various air quality conditions, ensuring a comprehensive assessment of indoor air quality. The performance of the DIH sensor within the HAM system was evaluated through comparison with brightfield microscopy, showing high concordance in size measurements. The efficacy of the DIH sensor was also demonstrated in two two-hour experiments under different environments simulating practical conditions with one involving distinct PM-generating events. These tests highlighted the HAM system's advanced capability to differentiate PM events from background noise and its exceptional sensitivity to irregular, large-sized PMs of low concentration.

**Keywords:** *Indoor Air, Particulate Matter, Digital Holography, Deep Learning, Air-quality Monitor*

# 1. Introduction

The COVID-19 pandemic has markedly heightened public awareness regarding particulate matters (PMs), underscoring their significance as a key indicator for indoor air quality assessment [1]. Notably, the monitoring of PMs is critical in contexts such as disease transmission prevention [2], the management of conditions such as asthma and allergies [3], and the maintenance of cleanroom standards in semiconductor fabrication [4]. The bulk of research in PM monitoring and analysis has traditionally concentrated on PMs less than 10 μm due to their sustained airborne presence and subsequent health implications [5, 6]. However, as shown in recent studies [7, 8], under ventilated conditions, PMs larger than 10 μm and beyond 100 μm can remain suspended and traverse similar distances as their smaller counterparts. Moreover, large PMs exhibit considerable variability in both morphology and composition, a diversity that can significantly affect their airborne suspension times beyond what is typically predicted for spherical particles [9]. This variance may result in prolonged air suspension, diverging from the expectations set by spherical models. Additionally, these large PMs possess increased surface area and volume, facilitating enhanced pathogen carriage [7], which, in turn, could amplify transmission and infection risks [10]. However, prevalent PM monitoring techniques predominantly target particles smaller than 10 μm, revealing a gap in the accurate detection of larger PMs [11]. While certain methodologies are applicable to larger PMs, they are not without their inherent drawbacks. Gravimetric analysis, for instance, offers reliability but is hindered by the necessity for prolonged material collection [12]. Techniques such as Laser Diffraction and Aerodynamic Particle Sizing (APS) encounter challenges in accommodating particles that deviate from spherical norms, thus restricting their effectiveness for larger PM assessment [13]. Furthermore, Shadowgraph imaging, despite its utility, is limited by its reliance on simplified particle geometries and the critical requirement for precise lens alignment, alongside a restricted depth of field, compromising its application [14].

Digital Inline Holography (DIH) has recently emerged as a cost-effective, and compact method for aerosol measurements [15-23]. This imaging-based approach employs a digital camera to capture interference patterns, or holograms, produced when light scattered by particles interferes with the un-scattered portion of the same coherent light source. DIH offers label-free characterization and operates effectively over a large depth of field, providing not just morphological data but also phase information, including 3D location and refractive index. Recent advancements in DIH for PM analysis have markedly improved the way we monitor air quality, offering innovative techniques for detecting and categorizing particles. For instance, Wu et al. introduced a DIH sensor that uses a lensless, impaction-based mechanism to achieve an impressive 93% accuracy in particle sizing [16]. This system directs particles onto an adhesive cover slip through a flow nozzle. However, its effectiveness is somewhat limited by the need for a 30-second impaction duration and cloud processing, which diminishes its self-sufficiency and necessitates regular cover slip replacements. To address this issue, Sauvageat et al. devised a real-time pollen monitor that leverages DIH to measure PM in a flow channel directly [17]. This device not only processes air rapidly at a rate of 40 liters per minute (LPM) but also employs advanced deep learning algorithms for accurate particle identification. Nonetheless, its utility is somewhat restricted to specific PM types (pollens in this case) due to its reliance on pre-concentration steps and an elliptical fitting model for machine learning. Furthermore, Kim et al. developed an innovative smartphone-based DIH sensor capable of estimating PM levels by capturing DIH patterns in a contained sample [23]. Notably, this sensor can process data in real time, using only the smartphone's built-in hardware. Despite its advancements, this model's inability to classify particles means it does not fully utilize the rich data that DIH measurements can offer.

Despite these advancements, all these existing PM monitoring methods based on DIH fall short of providing real-time surveillance with the desired throughput, specificity, and generalizability required for widespread indoor air quality applications. To bridge this gap, we propose the development of a sensor system dedicated to indoor air quality assessment, named Holographic Air-quality Monitoring (HAM). HAM is specialized in accurately identifying detailed information (e.g., concentration, size distribution, types) of common indoor PMs that are larger than 10 μm. Moreover, the system is equipped with an array of other sensors designed to capture comprehensive information of indoor air quality. HAM is engineered to match the high-throughput capabilities of leading non-imaging commercial particle analyzers, such as TSI APS devices and optical particle counters, setting a benchmark in the industry. The ultimate goal for HAM is to achieve sustained, uninterrupted, and autonomous operation for long-term indoor air quality management.

The following Methods section will outline the HAM system's design and operation, covering the hardware setup, DIH sensor technology, software design including general data processing pipeline and deep learning (DL) method for PM analysis. The Results section will present results related to the validation and demonstration of the HAM system, followed by a Conclusion and Discussion section.

## 2. Methods

The holographic air-quality monitor (HAM), as shown in Figure 1, comprises a hardware component that acquires data on PMs and other environmental conditions such as temperature, humidity, $CO_2$, and volatile organic compound (VOC) (Fig. 1a), and the corresponding software component that handles sensor control, data acquisition, analysis, and display. The hardware is paired with a GPU laptop (RTX 3070) that operates the software, while a graphical user interface (GUI) has been developed to disaply real-time sensor readings (Fig. 1b)

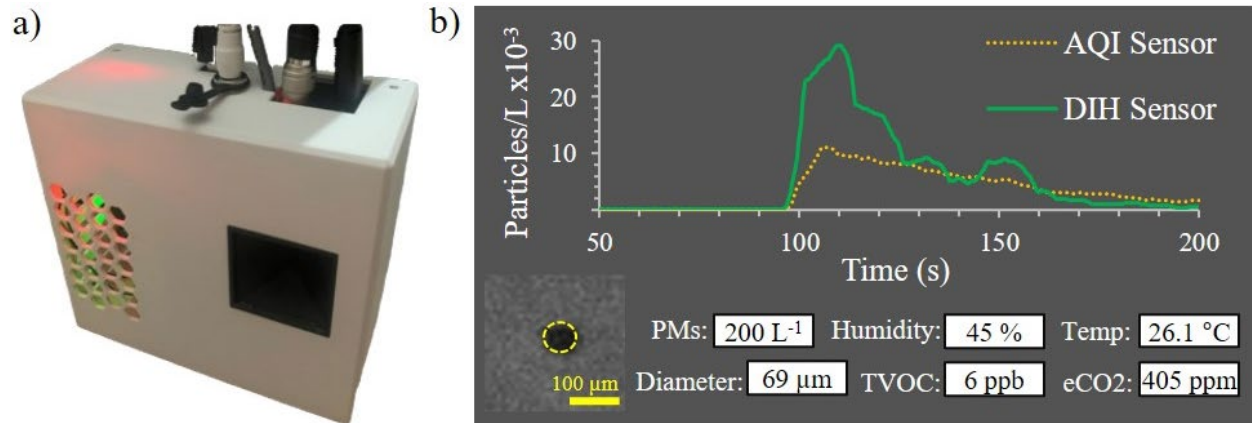

Figure 1: Overview of holographic air-quality monitor (HAM) including a) HAM system hardware housed in 3D printed enclosure and b) The graphical user interface (GUI) of HAM software displaying the sensor readings.

### 2.1. HAM Hardware

Central to the HAM hardware is the DIH sensor, designed for accurate measurement of large PMs ranging from 10 to over 300 μm, which operates in conjunction with a suite of environmental sensors for a comprehensive environmental assessment, as shown in Figure 2(a). These sensors include the PMSA003I sensor that uses laser diffraction to detect PMs as well as SGP30 and SCD-30 sensors, the combination of which provides temperature, humidity, effective $CO_2$, and total volatile organic compounds (TVOC). Specifically, PMSA0031 sensor (referred to as AQI sensor hereafter) provides direct readings for PM1.0, PM2.5, and PM10 and readings for PMs above 2.5 µm and 10 µm (referred to AQI2.5 and AQI10 hereafter), respectively, which are used to compare with our DIH sensor measurements later [24]. Coordination and control of these components are managed by an onboard microcontroller, an Arduino Mega2560. A USB 3.1 connection ensures high-speed image acquisition, while USB serial communication facilitates sensor operation, all powered via a 5V DC input.

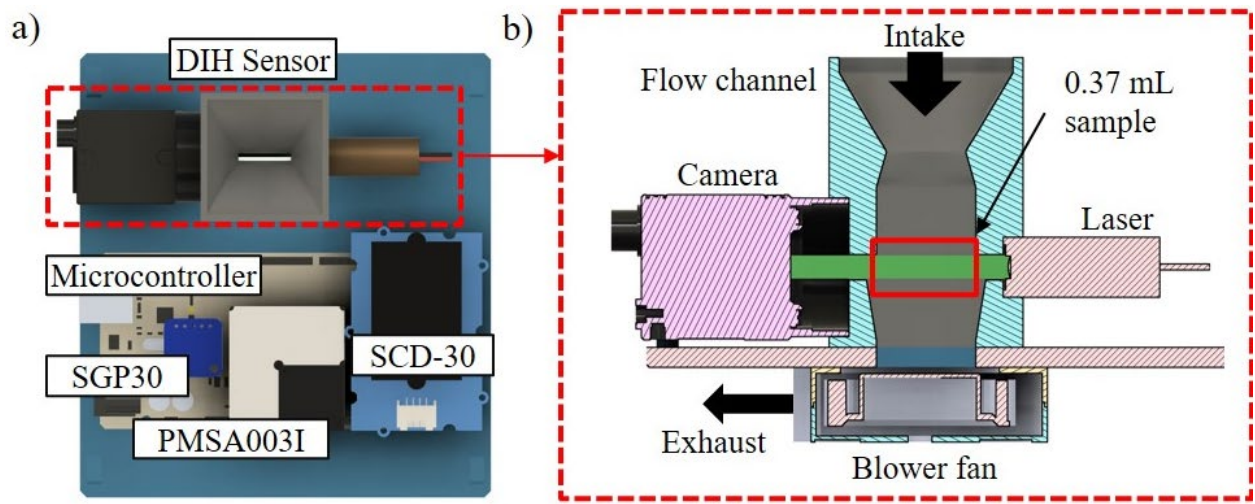

Figure 2: a) Sensors and layout of HAM system and b) cross section of DIH sensor showing the position of flow channel with respect to blower fan, laser, and camera, along with the direction of air flow.

The DIH sensor consists of a flow sampling unit with a converging nozzle, blue body in cross section of Figure 2(b), that channels air into a rectangular cross-section at a flow rate above 26 LPM. Air flow is generated using a constant pressure blower fan at the outlet of the nozzle calibrated using water displacement. The design of this nozzle is validated to ensure there is no sampling bias for PMs larger than 10 µm, with details included in the supplemental materials. The DIH sensor employs a lensless imaging design with a 520 nm (50 mW) pulsed laser and a global shutter camera (FLIR Blackfly S 1.6 MB camera) with a maximum frame rate of 226 frames per second (FPS) and resolution of 3.45 µm/px. The laser and camera are directly connected and sealed using the body of the flow sampling unit, Figure 2(b). The camera sensor, 5.0 mm × 3.7 mm, is matched to the sampling window, the full height and width of the smallest portion of the nozzle. A sample depth of 20 mm gives our final sample volume of 0.37 mL. During operation, the camera functions at 110 fps to acommendating the realtime processing speed of the GPU laptop. To enhance throughput and prevent image blurring, the laser emits 1210 evenly distributed pulses per second, each 629 ns wide, resulting in 11 exposures per image. These operational conditions ensure that the camera captures images quickly enough to accommodate the sensor's high sample flow rate of 28 LPM.

## 2.2. HAM Software

The HAM system includes a Graphical User Interface (GUI) with control panels that regulate the sampling flow rate and synchronize the operations of the onboard sensors. Specifically, for the DIH sensor, the software adjusts laser parameters, such as the number of pulses and pulse width, along with camera settings, including frame rate and gain. The software acquires and processes data from all sensors and displays the results in real time including particle counts and size distribution from both DIH and AQI sensors.

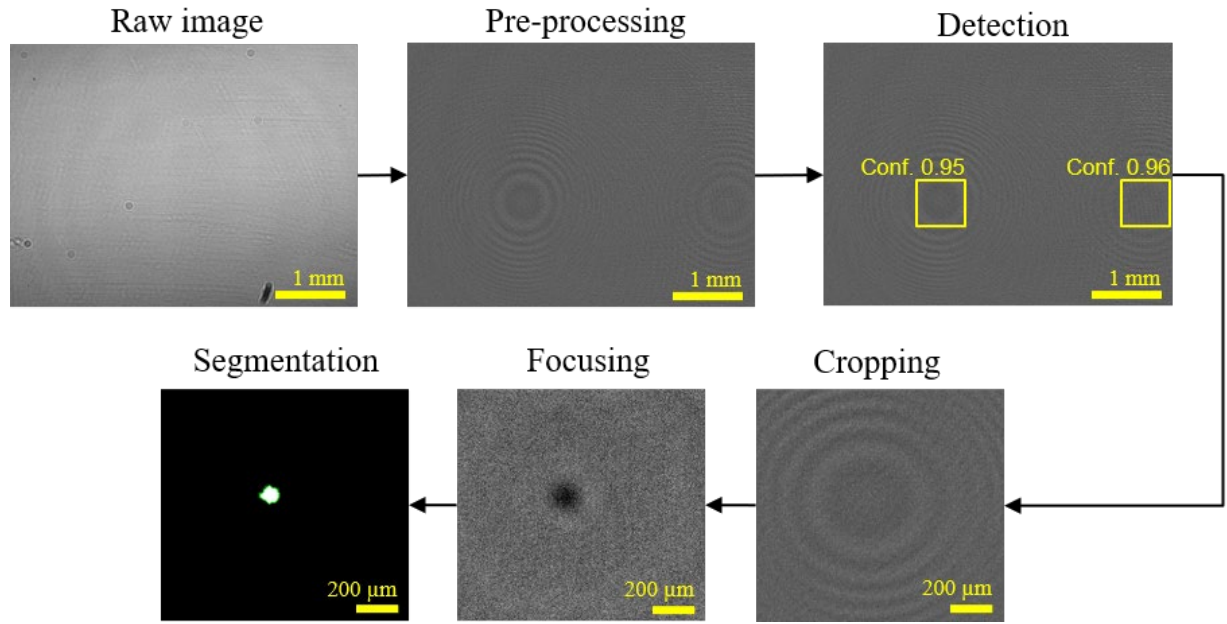

Figure 3: The hologram processing pipeline for the DIH sensor embedded in the HAM system including pre-processing, detection, cropping, focusing, and segmentation steps.

The DIH processing section of the HAM software performs PM detection and characterization through a multistage approach, as illustrated in Figure 3. This approach includes pre-processing, detection, focusing, and segmentation stages. In the pre-processing stage, raw images are enhanced by applying a moving window background subtraction with a window size of 40 frames. This process effectively removes noise, enabling the accurate detection of transient holographic signatures. The detection stage employs a Deep Learning (DL) model, which will be further discussed in the next paragraph, to accurately identify and isolate particle holograms (labeled using bounding boxes). Once a PM is detected, the hologram surrounding the detection center is cropped. The correct focal plane of the cropped hologram is then determined by reconstructing the holograms at various focal depths and applying the Tenengrad variance focus metric, which uses the variance of the Sobel gradient magnitudes across different depths to identify the in-focus image of the particle [25]. The in-focus particle is then segmented using a border-following algorithm to delineate its contour and extract information about its size, quantified as the area-based equivalent diameter, and its morphology [26]. Note that a 3-pixel Gaussian blur is applied before the segmentation approach to minimize the impact of noise on the segmentation process. The entire process operates in real time at 110 fps, with sizing up to 4000 counts/L using an RTX 3070 GPU .

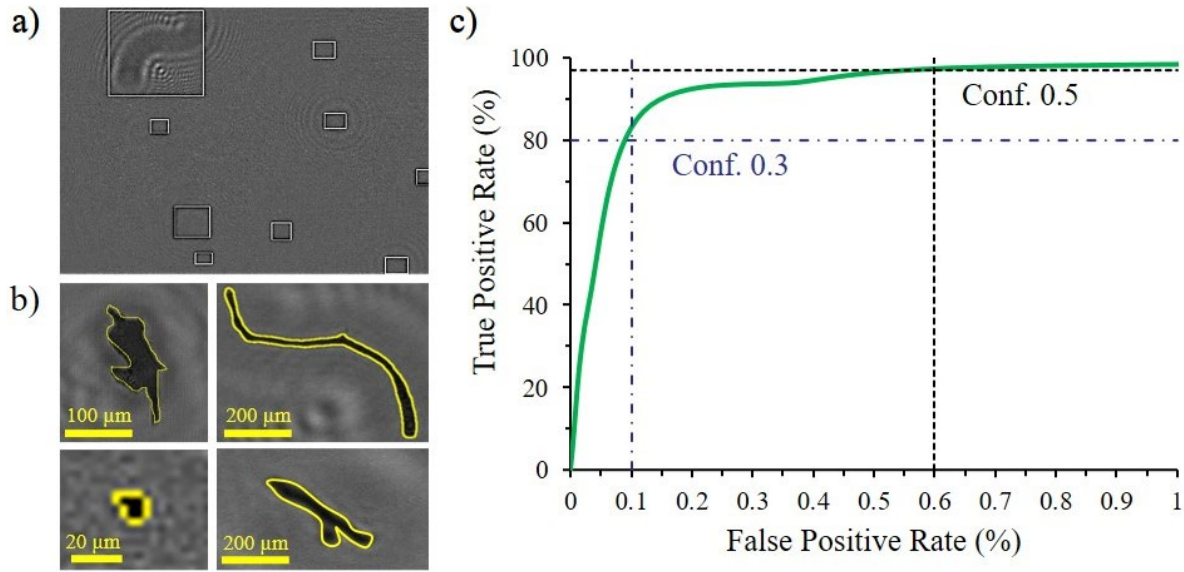

Figure 4: a) A sample hologram demostrating the detection (marked by white bounding boxes) of a variety of PMs using our deep learning model and b) A gallery of reconstructed (in-focus) images of detected PMs with different morphologies and sizes and c) receiver operator characteristic (ROC) curve of the deep learning model. Two operating points corresponding to confidence thresholds of 0.5 and 0.3 are illustrated in the ROC curve as the intersections of the pairs of vertical and horizontal dashed lines, leading to 97% true positive rate (TPR) and 0.6% false positive rate (FPR), and 80% TPR and 0.1% FPR, respectively.

The DL Detection Model uses the YOLOv5 framework, tailored for holographic PM detection through a customized training procedure [27]. Similar deep learning framework has been boardly applied to various realtime particle detection tasks using digital holography recently [28-31]. Initially, the model underwent pre-training on a diverse, manually annotated experimental dataset comprising approximately 4,000 instances of various particles, including yeast, plankton, E. coli, Campylobacter, and water spray droplets, covering a broad range of sizes and shapes. Subsequently, the primary training phase utilized a synthetic dataset, derived from 200 distinct in-focus PMs captured by the HAM system, illustrated in Figure 4(b). Specifically, this synthetic dataset consists of 2,000 synthesized particle holograms. Each image is generated through a blending process which places 5 to 10 PM holograms at randomized locations, scales and orientations using alpha blending. The PM holograms were generated by reconstructing in-focus PMs to a random imaging depth. The selected particle concentration for our synthetic holograms corresponding to about 4000 particles/L, matching the upper operational limit of our current HAM system for real-time particle analysis. To further enhance the diversity of the training dataset, the synthetic holograms are augmented using Gaussian blur, adding ISO noise, adjusting contrast and brightness, resulting in a final augmented dataset of 4,000 synthetic holograms of particles. Training was done over 10 iterations, lasting 100 epochs each, included re-augmentation between each iteration, to introduce novel variations while incorporating hard data mining to focus on the most challenging particles for the model. The final model, trained on 40,000 objects, was validated on a set of 1,000 manually labeled experimental particles. The model performance was evaluated using a receiver operator characteristic (ROC) curve, as illustrated in Figure 4(c). True positive rate (TPR) is defined as a correct prediction of a positive outcome, and the false positive rate is defined as an incorrect prediciton of a positive outcome. The figure also shows that, at a confidence threshold of 0.5, the model exhibits exemplary detection capabilities, achieving a True Positive Rate (TPR) of 97% and a False Positive Rate (FPR) of 0.6%, demonstrating the potential of our deep learning approach in accurately detecting PMs in holograms and setting a foundational basis that invites further application-specific validation.

## 3. Results

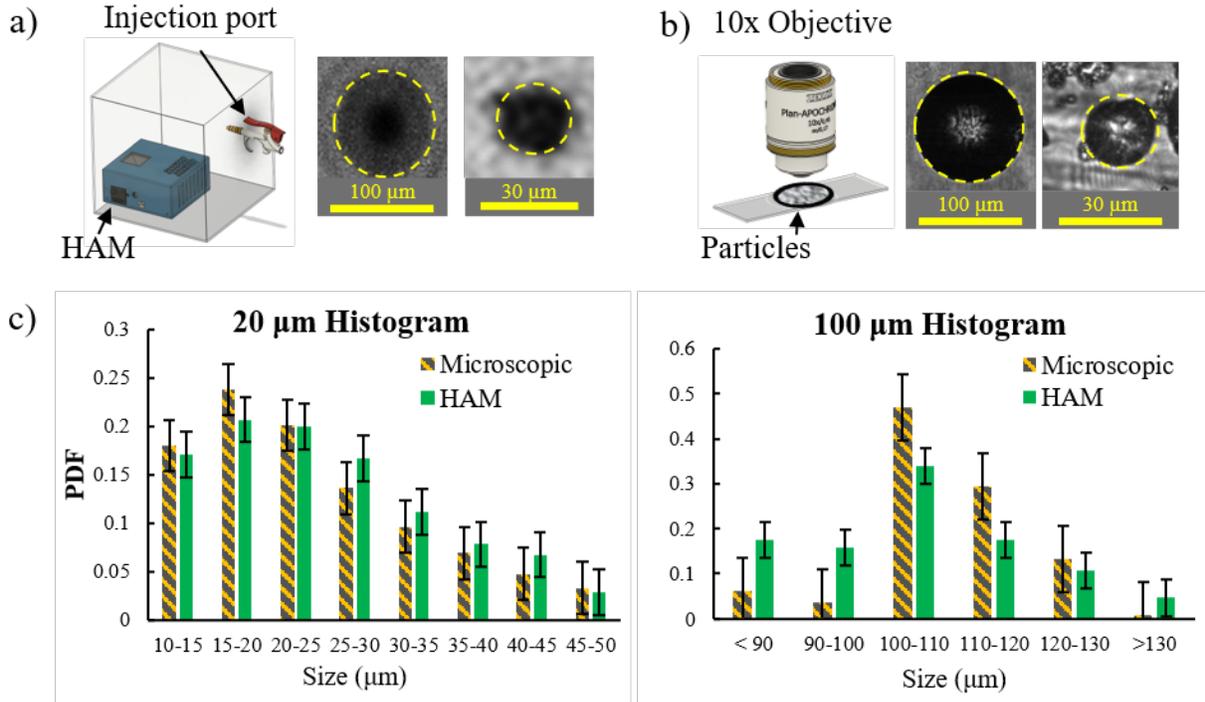

Figure 5: Comparison of the PM measurements obtained using HAM with those from conventional brightfield microscopy. Illustration of the experimental setup and sample results for a) HAM system and b) brightfield microscopy experiments using the same sets of particles. Size histograms of c) Size distributions of 20 μm silver-coated glass particles and 100 μm PMMA particles measured using both brightfield microscopy (yellow stripped) and our HAM system (green). Error bars represent the standard deviation.

### 3.1. Sizing performance validation using standard test particles

To evaluate the accuracy of our PM measurement using the HAM system, a validation experiment was conducted comparing HAM with brightfield microscopy for sizing 20 μm silver coated glass particles and 100 μm PMMA (polymethyl methacrylate) spheres. For this test, the HAM system was placed in the bottom center of a small-sealed enclosure, 22 cm × 22 cm × 22 cm where 100 mg of each particle type was introduced through an injection port while the HAM system is, actively running, shown in Figure 5(a). Compressed air, aided by a bypass for particle loading, was used to inject the particles into the enclosure. The encloure was thoroughly cleaned between the two tests with different particles. A total of 1200 particles were sized for each particle type. For brightfield microscopy, around 500 particles of each type were placed on a glass slide and sized using a 10x magnification, as shown in Figure 5(b). The microscopic images were analyzed using watershed segmentation for size distribution measurements. Figure 5(a) and Figure 5(b) show examples of the in-focus particles from DIH sensor (after reconstruction) and brightfield microscopic images of particles, providing a qualitative comparison that highlights the similarity between these two measurements.

Figure 5(c) presents size distribution histograms for 20 μm and 100 μm particles using 5 and 10 um bin size, respectively. The error bars shown, display the standard error of the mean of each size bin. The size distribution estimated from the HAM system closely align with that measured using bright field miscropy for both sizes. The histograms demonstrate compatibility within the margin of error for both sizes, with the 20 μm particles showing a notably close alignment. Note that the 100 μm particles exhibit a broader sizing variance, indicative of systematic discrepancies rather than a size-specific bias, possibly due to difficulties in accurately defining the boundaries of transparent PMMA particles. This issue, likely stemming from limitations in the focusing metric for transparent materials employed in our current hologram processing pipeline, is exemplified in the sample image of a 100 μm particle in Figure 5(a).

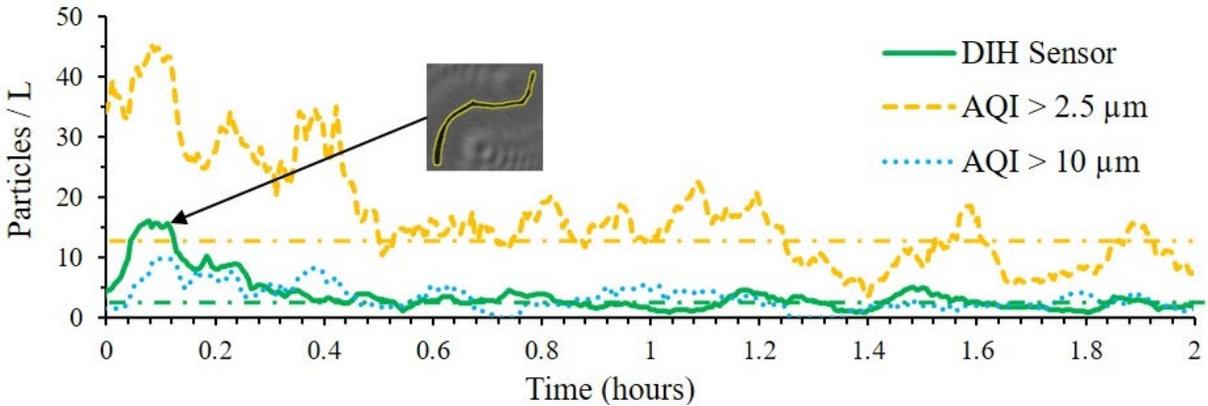

Figure 6: The comparison of PM concentration measurements from the DIH and AQI sensors embedded in the HAM system for two-hour experiment in a 3.6 m (length) × 3.6 m (width) × 2.8 m (height) living room.

## 3.2. Demonstration experiment for indoor air quality monitoring

Furthermore, we assessed the HAM system's noise floor and sensitivity for PM measurements under a simulated practical condition through a two-hour experiment in a carpeted living room. The system's sensors were in continuous operation, gathering data both prior to and subsequent to a singular perturbation induced by the act of shutting the room's blinds at time 0. This experiment was conducted in the center of the living room with all ventilation turned off to eliminate external airflow influence. The HAM system was placed directly in the center of the 3.6 m (length) × 3.6 m (width) × 2.8 m (height) room on a medium height polypropylene carpet. Ambient temperature and relative humidity were monitored and stable at 31±1 ºC and 25±4 %, respectively, during the two-hour experiment. Across the entire experiment, TVOC levels remained below 10 ppb and $CO_2$ levels of 1000±150 ppm. The blinds, polyester 2.0 m (width) × 1.2 m (height), were initially closed on the back center wall of the room, hanging on a rod 0.3 m below the ceiling. The perturbation was created by pulling the blinds from one side in a smooth continuous motion lasting approximately 1 s. Following the initial disturbance, the room was kept devoid of movement or occupants, ensuring an undisturbed environment for the HAM system to accurately assess the generation of PM associated with the initial disturbance and differentiate it from background noise.

The results, as depicted in Figure 6, show that both the DIH sensor and AQI2.5 successfully captured the trend associated with the decay of PMs following the disturbance. The decay time of the disturbance, as measured by both DIH and AQI2.5, aligns closely at approximately 30 minutes. However, a notable difference was observed in the noise floor levels, with the AQI2.5 sensor displaying a higher noise floor (12 particles/L) and larger fluctuations compared to that of the DIH (2 particles/L). The augmented noise associated with the AQI2.5 could be ascribed to the presence of a higher concentration of PMs below the DIH sensor's detection threshold and their prolonged suspension time in an environment devoid of ventilation. The DIH sensor revealed that a significant portion of the large PMs at the peak consisted of elongated fibers, likely originating from the carpet. In contrast, the AQI10 readings faced challenges in identifying the peak corresponding to the observed large, irregular particles. Furthermore, AQI10 readings displayed more pronounced fluctuations relative to the DIH measurements. This variation suggests that although the AQI10 sensor can offer a general indication of particulate levels, its precision is notably affected by substantial noise, particularly in contexts involving larger or irregularly shaped particles.

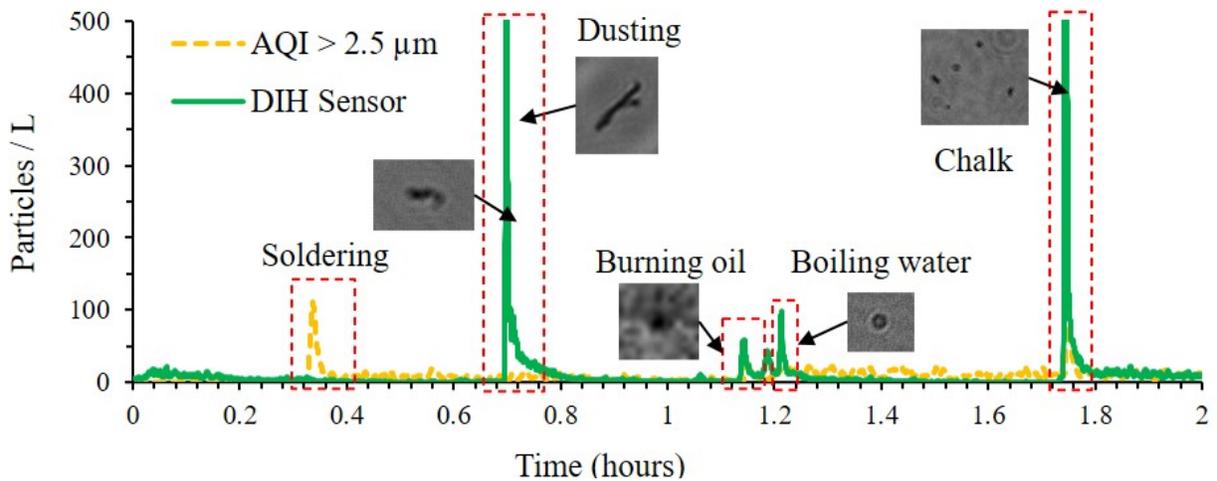

Figure 7: The comparison of PM concentration measurements from the DIH and AQI sensors embedded in the HAM system for two-hour experiment in a 1.2 m (length) × 1.0 m (width) × 2.8 m (height) room under various PM generation events.

To further assess the HAM system's ability to detect and monitor various PM generation events, another two-hour deployment was conducted in a more compact, ceramic tiled, room with dimensions of 1.2 m (length) × 1.0 m (width) × 2.8 m (height). The HAM system was placed on the countertop against the long wall of the room, at a height of 1 m from the floor, approximately in the center of the counter. The temperature and relative humidity at the beginning of the experiment were 31.6 ºC and 21.7 %, respectively. $CO_2$ and TVOC initial concentrations were 800 ppm and 10 ppb, respectively. This environment, with all ventilation turned off, allowed for easy observation of distinct PM events. The HAM system was positioned approximately 0.5 m away from where the PM generation events were introduced. Throughout this experiment, all sensors within the HAM system operated continuously, ensuring a comprehensive evaluation of its performance under simulated practical conditions.

Five distinct PM generation events were introduced to assess the HAM system's capabilities. The room was allowed time to settle below 20 particles/L before each new event. One notable exception here are the burning oil and boiling water events, which directly precede each other, this was done to simulate a "cooking" event. The initial event, soldering, involved the generation of smoke through the melting of solder, commencing at the 20-minute mark and continuing for roughly 5 minutes. This was followed by dusting cabinets and surfaces using a damp hand cloth, initiated 40 minutes into the experiment and lasting 2 minutes. The subsequent event involved heating oil on an electric burner until it reached it began to smoke, heating began at 60 minutes, with smoke appearing around 67 minutes, after which heating was ceased immediately. Figure 8, showing our onboard sensor data, we notice the room temperature increase from the baseline of about 32 ºC, to 36 ºC over 20 minutes, beginning at 60 minutes, along with peaks in relative humidity, $CO_2$ and TVOC. $CO_2$ did not return to its baseline value, instead spiking to 1350 ppm, then stabilizing at 1250 ppm. Boiling water constituted the next event, with heating beginning at 70 minutes, reaching boiling point by about 72 minutes, and then promptly discontinued. The final event involved the dispersion of 10 grams of fine powdered chalk into the environment, occurring 1 hour and 45 minutes into the experiment. Beginning shortly after this dispersal of chalk dust, at approximately 1.8 hours, we again see a rapid increase in temperature, relative humidity, $CO_2$, and TVOC.

Figure 7 presents the variation of PM concentration measured using both DIH sensor and AQI2.5 sensor embedded in the HAM system during the two-hour experiment with dashed boxes demarcating the time spans corresponding to the five introduced PM generation events. Remarkably, the DIH sensor detected PM emissions from four out of the five events—dusting, oil burning, water boiling, and chalk dust generation—each characterized by pronounced, sharp peaks that markedly surpassed the ambient noise level of approximately 2 particles/L. These results indicate the DIH's high sensitivity and its ability to differentiate between background noise and PM from specific activities. Moreover, the DIH sensor captured particles with unique morphologies corresponding to each of the four activities, illustrating its capability to discern and characterize the distinct physical properties of PMs. Instances of these in-focus particles are depicted in Figure 7. Specifically, during the dusting event, elongated fibers ranging between 50 to 300 μm were detected. The oil burning episode produced oil droplets measuring approximately 7.5 to 10 μm, whereas the boiling water event yielded water droplets sized between 10 – 20 μm. Although these events were initiated one after the other, our sensor had sufficient temporal resolution to separate the peaks, with classification of the peaks done through manual identification of the particles as either oil or water. The generation of chalk dust unveiled a diverse array of shapes and sizes spanning 7.5 to 100 μm. According to the literature, water and oil droplets are typically found in the size range of 1-1000 μm, and chalk particles in the range of 5-100 μm [32, 33], corroborating the observations of the DIH sensor.

In comparison, among the four events detected by DIH sensor, the AQI2.5 sensor identified PMs from oil burning and chalk dusting yet failed to register significant signals from the dusting and boiling water events. This discrepancy can be ascribed to its design limitations, which are not conducive to detecting irregularly shaped and larger particles, leading to an absence of signal for the dusting event and a minimal signal for the boiling water, due to the prevalence of particles exceeding 7.5 μm. Nonetheless, it adeptly recorded PMs from burning oil and chalk dusting, attributed to the smaller dimensions of these particles and the diverse shapes found in the chalk dust, though the spikes were less pronounced than those detected by the DIH. Notably, the DIH did not detect PMs from the soldering activity, whereas the AQI2.5 sensor exhibited a distinct spike, presumably due to the soldering smoke comprising very fine particles (< 1 μm) that fall below the DIH sensor's operational size range. This variation in sensor efficacy underscores the complementary strengths of the DIH and other sensors in tracking an extensive array of PM sizes and forms, with the DIH demonstrating enhanced sensitivity and specificity for larger and more uniquely shaped particles.

# 4. Conclusions and Discussion

In this research, we introduce a novel holographic air-quality monitor (HAM) system, specifically designed for the detailed monitoring of large particulate matter (PM) with diameters greater than 10 μm, which are of significant concern for public health. At the heart of the HAM system lies lensless digital inline holography (DIH) sensor, composed of a pulse laser, a digital camera, and a flow sampling channel. The sensor operates with a deep learning model to achieve the real-time detection and analysis of PMs, including their size and morphology, as they pass through the sampling channel with an airflow rate of 26 liters per minute (LPM). Complementing this functionality, the system includes sensors for the measurement of smaller PMs, humidity, temperature, $CO_2$ levels, and Total Volatile Organic Compounds (TVOC), thereby providing comprehensive assessment of indoor air quality. The PM measurement performance of the DIH sensor embedded in the HAM system was assessed through a comparative analysis with brightfield microscopy, exhibiting a high degree of concordance. The proficiency of the DIH sensor was further showcased in two-hour experiments within both a 3.6 m × 3.6 m carpeted room and a smaller, 1.2 m × 1.0 m tiled space, simulating practical conditions with five distinct PM-generating events. These tests underscored the HAM's superior performance in differentiating PM events from background noise, exceeding that of conventional sensors, and its acute sensitivity to irregular and large-sized, low-concentration PMs that often elude the detection range and assumptions of standard PM monitoring tools.

The HAM system represents a significant advancement in air quality monitoring, distinguished by its capability to detect and analyze large PMs that are often overlooked by conventional sensors. This innovation addresses a critical gap in indoor air quality assessment by providing detailed, assumption-free insights into various large particles through advanced holographic imaging, which captures both the optical and morphological properties of particles. Unlike traditional methods, the HAM system not only counts particles but also enables real-time detection and analysis of potential airborne hazards, including pollen, fungi, mold spores, asbestos fibers, microplastics, and dust mites. By accurately characterizing these significant threats, the HAM system plays a crucial role in assessing risks associated with respiratory illnesses, allergies, and other health conditions linked to poor air quality. Furthermore, the system's deep learning approach enhances its ability to identify unknown particles and emerging hazards by detecting anomalies and outliers in PM attributes. Operating at a high throughput rate of 26 LPM, the HAM system surpasses traditional imaging-based methods and rivals lower-fidelity sensors, ensuring comprehensive and accurate air quality data. With its cost-effective, modular, and compact design, the HAM system can be seamlessly integrated into existing or new sensor networks, significantly improving their ability to monitor and characterize large PMs. This enhancement is a pivotal step toward developing new strategies for mitigating, preventing, and controlling airborne health and economic hazards, underscoring the HAM system's substantial contributions to public health and environmental safety.

Beyond indoor air quality, the development of the HAM system holds substantial potential across various other fields. In climate and atmospheric science, the system's ability to monitor coarse mode aerosols, such as cloud droplets, icelets, desert dust, and sea salt, is critical for understanding the Earth's radiative balance and cloud formation and precipitation processes. The HAM system's advanced particle characterization capabilities can also be applied in environmental science to monitor pollutants in water bodies, in industrial settings for detecting hazardous dust and fibers, and in agricultural contexts for tracking pollen distribution and plant pathogens. Its adaptability to different media, combined with high precision and real-time analysis, makes the

HAM system a versatile tool in diverse applications where understanding particulate matter is essential. This versatility underscores the broader impact of the HAM system, paving the way for innovations in environmental monitoring, public health, climate science, and beyond.

It should be acknowledged that while the HAM system delivers consistent real-time PM analysis at a 26 LPM sampling rate, it is optimized for environments where particle concentrations do not exceed 4000 particles/L. This threshold is well within the bounds of typical indoor settings, where particulate matter larger than 5 μm seldom surpasses 1000 particles/L, ensuring the HAM's broad applicability for indoor air quality evaluation [34]. Although our current investigation primarily establishes a proof of concept, forthcoming efforts will be directed towards enhancing the HAM system with cloud-based data processing, streamlining the analysis workflow, and reducing the system's footprint and cost to bolster reliability. These advancements will pave the way for distributed sensor networks, offering a more expansive monitoring of indoor air quality.

**Data Avaialbility**:

The data that support the findings of this study are openly available in GitHub at https://github.umn.edu/HongFlowFieldImagingLab/HAM.

**Funding Statement:**

No funding was received for this research.

**Supplementary Material:**

The supporting material, "Holographic Air-quality Monitor (HAM) Supporting.docx", included with this study overviews the design and valdiation of the HAM systems air sampling nozzle, providing information on the sampling bias and validation methods.